# C-Arm Non-Circular Orbits: Geometric Calibration, Image Quality, and Avoidance of Metal Artifacts

P. Wu,[1] N. Sheth,[1] A. Sisniega,[1] T. Wang,[1] A. Uneri,[1] R. Han,[1] R. Vijayan,[1] P. Vagdargi,[1] B. Kreher,[2] H. Kunze,[2] G. Kleinszig,[2] S. Vogt,[2] S.-F. Lo,[3] N. Theodore,[3] and J. H. Siewerdsen[1,3]

[1]Department of Biomedical Engineering, Johns Hopkins University, Baltimore MD USA
[2]Siemens Healthineers, Erlangen Germany
[3] Department of Neurosurgery, Johns Hopkins University, Baltimore MD USA

## ABSTRACT

Metal artifacts present a frequent challenge to cone-beam CT (CBCT) in image-guided surgery, obscuring visualization of metal instruments and adjacent anatomy. Recent advances in mobile C-arm systems have enabled 3D imaging capacity with non-circular orbits. We extend a previously proposed metal artifacts avoidance (MAA) method to reduce the influence of metal artifacts by prospectively defining a non-circular orbit that avoids metal-induced biases in projection domain.

Accurate geometric calibration is an important challenge to accurate 3D image reconstruction for such orbits. We investigate the performance of interpolation-based calibration from a library of co-registered circular orbits to estimate the projection matrix for any non-circular orbit. We apply the method to non-circular scans acquired for MAA, which involves: (i) coarse 3D localization of metal objects via two scout views (without strong priors on the patient or implants) using an end-to-end trained neural network; (ii) calculation of the metal-induced x-ray spectral shift for all possible views; and (iii) identification of the non-circular orbit that minimizes the variations in spectral shift.

Non-circular orbits with interpolation-based geometric calibration yielded reasonably accurate 3D image reconstruction with a slight loss of spatial resolution [frequency at 10% MTF ($f_{10}$) reduced by 12%]. Non-circular scans demonstrated expected benefits to 3D sampling – e.g., a ~40% reduction in cone-beam artifact in a Defrise disk phantom. The end-to-end neural network accurately localized metal implants with just two scout views even in complex anatomical scenes, improving Dice coefficient by ~42% compared to a more conventional cascade of separately trained U-nets. In a spine phantom with pedicle screw instrumentation, non-circular orbits identified by the MAA method reduced the magnitude of metal "blooming" artifacts (apparent width of the screw shaft) in CBCT reconstructions by ~70%.

The proposed imaging and calibration methods present a practical means to improve image quality in mobile C-arm CBCT by identifying non-circular scan protocols that improve sampling and reduce metal-induced biases in the projection data.

## I. INTRODUCTION

Intraoperative cone-beam CT (CBCT) is increasingly used for guidance and validation in placement of surgical instruments [1], [2], including spinal pedicle screws, neurosurgical electrodes etc. Such highly attenuating metal objects within the field-of-view (FOV) can cause metal artifacts that degrade image quality and confound visualization of adjacent anatomical structures. Many algorithms have been proposed for metal artifacts reduction, including classic projection / image domain postprocessing ("MAR") methods [3] and polyenergetic reconstruction methods.[4] However, these methods contend with artifacts *after* biases are already incurred in the projection data. A recent method (called metal artifact *avoidance*, MAA) [5] proposes to acquire projection data with minimal metal-induced bias, allowing MAR or polyenergetic reconstruction to work even better – or in some cases yielding "raw" CBCT images with artifacts reduced to a level sufficient for the imaging task.

Increasingly, interventional CBCT systems are capable of computer-controlled motion of the x-ray source and detector for general, non-circular orbits that can e.g. improve sampling characteristics,[6] and maximize the spatial-frequency-dependent signal and noise transfer characteristics with respect to the imaging task.[7] Such orbits also provide a means to reduce the influence of metal instrumentation (i.e., reduce / avoid metal artifacts) by collecting data in a manner that minimizes the factors of metal-induced bias (e.g. beam hardening). Recently, such capability for complex source-detector orbits is available on mobile C-arms with motorized control of multiple axes of gantry motion.

Compared to fixed-room, robotic C-arms, such mobile systems feature a relatively small footprint, lower cost, and flexibility in workflow, bringing the potential for non-circular orbits to surgical guidance.

In this work, we report the capacity for 3D imaging with non-circular orbits on a mobile C-arm with motorized tilt and angulation of the gantry. Specifically, we address issues of geometric calibration for general (not pre-defined) non-circular orbits and evaluate 3D imaging performance characteristics of non-circular orbits compared to conventional circular orbits. We use such capability to extend the MAA method [5] to non-circular orbits that minimize metal-induced bias. The method is shown to operate well with just two low-dose scout views (without other prior information of patient anatomy / metal objects) and is compatible with MAR and polyenergetic reconstruction methods that can further improve image quality.

## II. METHODS

### A. Non-Circular Orbits on a Mobile C-Arm System

The mobile C-arm used in this work (Cios Spin 3D, Siemens Healthineers, Forcheim, Germany) is shown in Fig. 1(a). The C-arm has motorized control of rotation angle ($\theta$: 0° to 196°) and gantry tilt ($\phi$: -30° to 30°, due to realistic considerations of patient / table collision), permitting non-circular orbits to be executed by variation of $\theta$ and $\phi$ during the scan. Tilted circular orbits (constant $\phi$) can be calibrated using established methods – e.g., Cho et al.[8] However, for non-circular scan trajectories defined by methods like "task-driven" imaging [7] and MAA, the orbit is designed on the fly and cannot be pre-calibrated given the large variety of feasible ($\theta$, $\phi$) combinations. We therefore developed an interpolation-based approach to address this issue, detailed as follows.

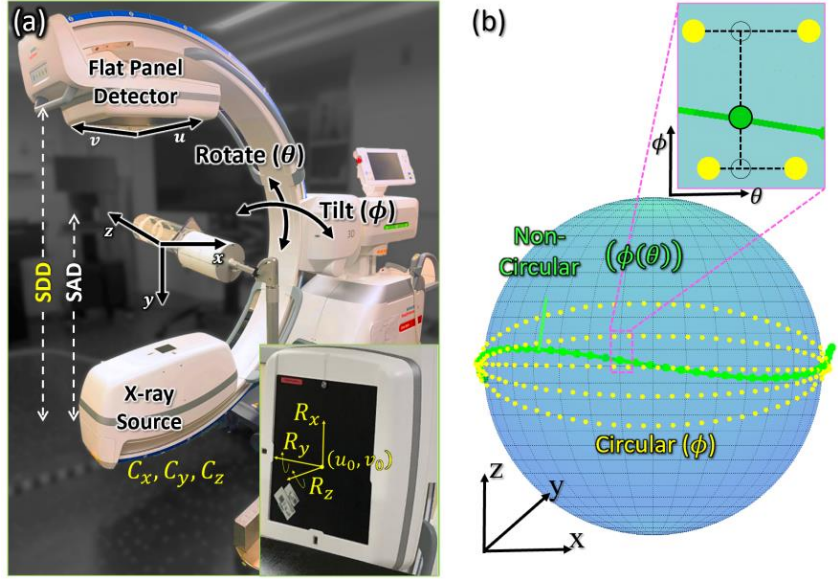

Fig. 1. Mobile C-arm used in this work (Cios Spin 3D, Siemens Healthineers). (a) C-arm geometric parameters and zoomed-in view of the detector (bottom right). The nine parameters (degrees-of-freedom) determined via geometric calibration are marked in yellow. (b) Source-detector orbits illustrated on a sphere. Circular orbits are in shown yellow (various settings of gantry tilt, $\phi$). An example non-circular orbit is shown in green ($\phi$ changing with $\theta$). The zoomed inset illustrates estimation of geometric parameterization for one view (green dot) determined by interpolation of four nearest vertices in a library of geometric calibrations of circular scans (yellow dots).

First, we built a library of geometric calibrations for circular obits with tilt ranging from $\phi$ = -30° to +30° at 5° intervals using a cylindrical BB phantom placed at the same tilt angle as the C-arm gantry.[8] Affine transformation of the library registers all calibrations into a common coordinate system (co-registration):

$$\boldsymbol{P}_{\phi=0} = \boldsymbol{P}_{\phi=\phi^*} \times T_{tilt} \times T_w \qquad (1)$$

$$T_{tilt} = \begin{bmatrix} 1 & 0 & 0 & 0 \\ 0 & \cos(\phi^*) & -\sin(\phi^*) & 0 \\ 0 & \sin(\phi^*) & \cos(\phi^*) & 0 \\ 0 & 0 & 0 & 1 \end{bmatrix} \quad T_w = \begin{bmatrix} 1 & 0 & 0 & t_x \\ 0 & 1 & 0 & t_y \\ 0 & 0 & r_z & t_z \\ 0 & 0 & 0 & 1 \end{bmatrix} \qquad (2)$$

where $\mathbf{P}_{\phi=\phi^*}$ is the projection matrix at tilt angle $\phi^*$, and $T_{tilt}$ is a transformation through tilt angle $\phi^*$ (read from the C-arm motor encoder). The transformation $T_w$ accounts for small discrepancies in displacement ($t$) and tilt ($r_z$) between the BB phantom and the gantry in each member of the library, determined by 3D-3D registration. This co-registration step: (i) makes sure that the orientation of the reconstructed image remains unchanged regardless of the orbit; (ii) improves the accuracy of the interpolation step below.

The co-registered calibrations were then decomposed into nine parameters: SDD (source detector distance), $C$ (source location, in $x, y, z$), $u_0, v_0$ (piercing point), and $R$ (detector rotation matrix, in $x, y, z$) as shown in yellow in Fig. 1(a). The system geometry for a general non-circular orbit can then be estimated by interpolating the geometric parameters from the calibration library – viz., the four nearest views from calibrated tilted circular orbits as illustrated in Fig. 1(b). Parameters SDD, $u_0$ and $v_0$ were estimated by scattered linear interpolation, and $C$, $R$ by spherical linear interpolation (slerp).[9] The slerp operation is non-commutative and was performed first in $\theta$ and then in $\phi$ for reduced interpolation error, since the $\theta$ direction is more finely sampled in the calibration library.

### B. Source-Detector Orbits for Metal Artifact Avoidance (MAA)

A flowchart representation of the MAA algorithm is shown in Fig. 2. The main stages (marked in blue squares) include: (i) coarse localization of metal objects from scout views; (ii) predictive calculation of metal-induced bias for all possible views; and (iii) non-circular orbit optimization. These stages are described in details as follows:

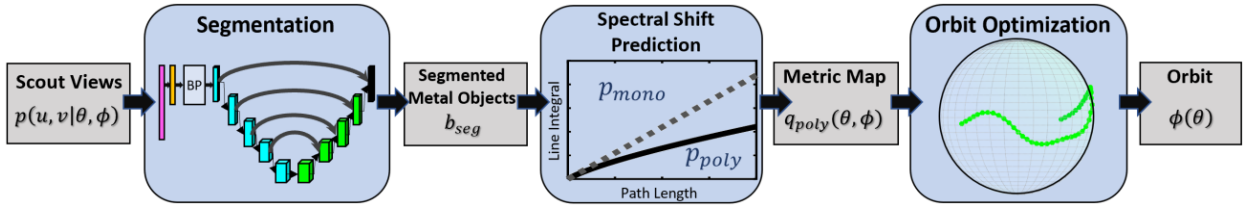

Fig. 2. Flowchart of the MAA algorithm. Two scout views are used for coarse 3D localization of metal objects ($b_{seg}$) using an end-to-end CNN. Biases arising from x-ray spectral shift are predicted as a function of gantry rotation angle and tilt. A non-circular orbit that best avoids metal-induced biases is determined by finding a curve in $q_{poly}(\theta, \phi)$ with minimal standard deviation in spectral shift.

#### 1) 3D Localization of Metal Objects from Two Scout Views

Given two scout views, the MAA method determines a binary 3D reconstruction (coarse segmentation, $b_{seg}$) describing the 3D location of metal objects in the FOV. Unlike in MAR, a coarse segmentation was found to be sufficient for MAA, requiring only an approximation of the shape, size, and orientation of metal objects (cf. MAR methods that typically require very accurate segmentation to avoid introducing secondary artifacts [10]).

Coarse segmentation used a novel, end-to-end convolutional neural network (CNN) shown in Fig. 3, which learns features simultaneously in projection and image domains. Specifically, projection domain feature extraction layers are connected to image domain 3D segmentation layers through a frozen (no learnable parameters) backprojection layer. Intuition underlying such a network is that image domain segmentation is improved by fusing shape information from the projection domain (before being smeared by backprojection), and projection domain feature extraction is guided by the segmentation loss backpropagated from the image domain.

Each projection domain feature extraction layer contains two 3×3 convolutions, each followed by a rectified linear unit (ReLU) with batch normalization, and in the end a 2×2 max pooling layer. The 3D segmentation layers were implemented as a three-layer multi-channel 3D U-Net (slightly modified from [11], detailed in Fig. 3) with general Dice coefficient as loss function. At the beginning of the 3D segmentation layers, the derivative of the loss function with respect to the input of the backprojection layer is the forward projection operation (determined with C-arm geometry and Siddon ray tracing,[12] not taken as learnable parameter of the network), enabling joint learning in projection and image domains. The network was "He normal" initialized and trained using the Adam optimizer with an initial learning rate of $5\times10^{-4}$ for 100 epochs.

One of the key hyper-parameters of the network is the number of feature maps extracted in the projection domain (denoted as $N_c$ as shown in Fig. 3), which is also the number of input channels for the multi-channel 3D U-Net in the image domain. The backprojection step (within the backprojection layer) is performed individually for each channel. As shown in studies below, having more than one feature maps is crucial to the performance of the proposed end-to-end network.

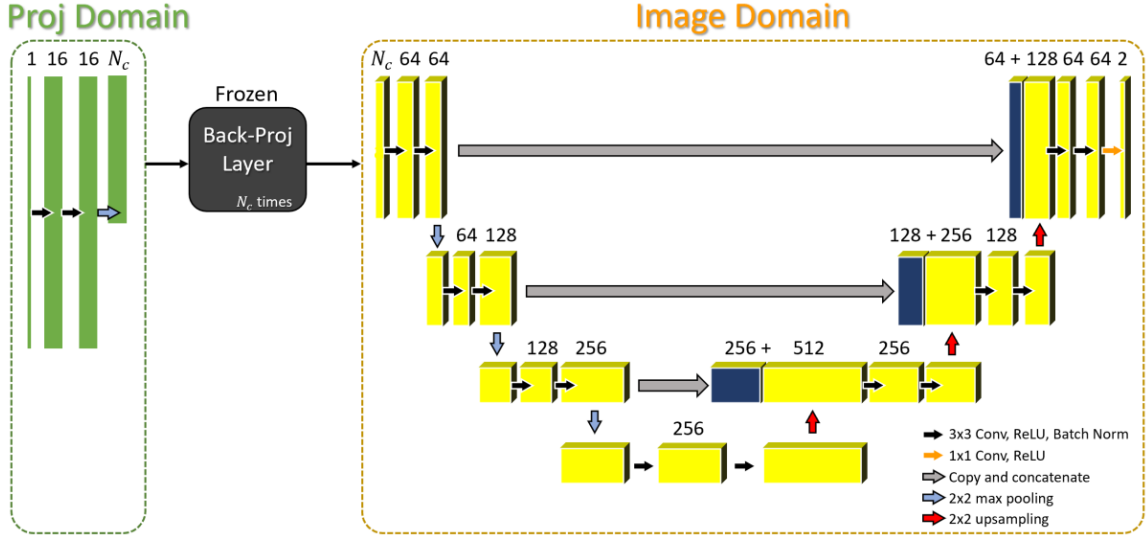

Fig. 3. End-to-end neural network for 3D localization / segmentation of metal instrumentation. Projection domain feature extraction layers are connected to the image domain segmentation layers through a frozen backprojection layer (determined from C-arm geometry), enabling joint learning in projection and image domains.

For simplicity and to avoid the requirement for vendor-specific metal instrument models, the network was trained with only simulated data, which was generated following the flowchart in Fig. 4. Digitally reconstructed radiographs (DRRs) were computed from 10 abdominal and thoracic CT images drawn from the Cancer Imaging Archive (TCIA) and a random number of generic metal objects: ellipsoids ranging in size and eccentricity (10 – 80 mm major and minor axes) and coarsely segmented spine screws drawn from CT images. Future work could certainly involve adding higher fidelity object models (e.g., vendor-specific designs) and other application-specific objects (e.g., fracture fixation plates) for improved performance. The effects of data truncation, beam hardening (through poly-energetic forward projection), scatter, lag, glare, and noise (quantum and electronic) were included in DRR generation. Data augmentation included variation in the location, size, orientation, aspect ratio, and attenuation of simulated metal objects in each DRR, resulting in a total of ~8,400 (8,000 training + 400 validation) images.

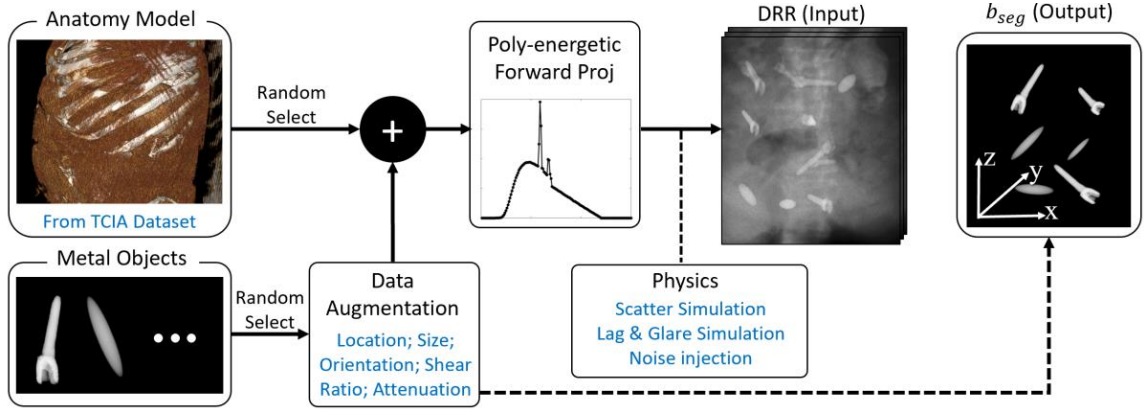

Fig. 4. Flowchart for training data generation.

### 2) Prediction of Metal-Induced Bias in Projection Data

The second step of the MAA method is to estimate the metal-induced bias (e.g., summation of x-ray spectral shift) in the projection domain for all rotation ($\theta$) and tilt ($\phi$) angles [5]:

$$q_{poly}(\theta, \phi) = \sum_{u,v} \left( p_{mono}(u,v|\theta,\phi) - p_{poly}(u,v|\theta,\phi) \right) \quad (3)$$

$$p_{mono}(u,v|\theta,\phi) = c(\mathbf{A} b_{seg}) \quad (4)$$

$$p_{poly}(u,v|\theta,\phi) = \sum_{i=1}^{N} \alpha_i p_{mono}^i(u,v|\theta,\phi) \quad (5)$$

where $q_{poly}$ is the spectral shift predicted for each $\theta$ and $\phi$, $p_{mono}(u,v|\theta,\phi)$ is the monoenergetic line integral, $p_{poly}(u,v|\theta,\phi)$ is the corresponding polyenergetic line integral affected by beam-hardening, **A** is the projection matrix, $c$ is an empirically estimated scalar representing the attenuation coefficient of the metal object, and $\alpha_i$ are precalculated polynomial mapping coefficients determined by the spectrum and metal material. The MAA method is relatively insensitive to both the spectral model and the metal material (including $c$), since the spectral model is only used to predict desirable orbits (not to perform an actual beam-hardening correction).

### 3) Orbit Optimization & Reconstruction

The third step of the MAA method is to identity the non-circular orbit that minimizes projection data biases (and thereby avoid metal artifacts in the 3D image reconstruction). Rather than simply summing the spectral shift at each $\theta$, we chose an objective that describes the *inconsistency* in spectral shift between projection views, since such inconsistencies are the underlying source of streaks commonly identified as metal artifacts. To reduce dimensionality and encourage a smooth non-circular orbit, we modeled $\phi(\theta)$ as a superposition of cubic b-splines:

$$\hat{f} = \arg\min_{f} = Q_{poly}(f) \tag{6}$$

$$Q_{poly}(f) = \sigma\left[q_{poly}\left(\theta, \sum_{i=0}^{M} f_i B(\theta - \theta_i)\right)\right] \tag{7}$$

where $Q_{poly}$ is a metric of spectral shift *variation* determined by the standard deviation operator ($\sigma$), $B$ is the cubic b-spline ($M=10$ control points), and $f_i$ is the control parameter for spline knot control points $i$. The non-convex optimization in Eqs. (6-7) was solved with the covariance matrix adaptation evolution strategy (CMA-ES) [13] solver operating in near real-time (~0.5 s runtime). The penalized weighted least squares (PWLS) method [14], [15] was used for image reconstruction for both circular and non-circular orbits.

## C. Experimental Study

### 1) Geometric Calibration of Non-Circular Orbits

The reproducibility of geometric parameters was evaluated for an example non-circular orbit ($\phi$ linearly increasing from -20° to +20° while $\theta$ linearly increasing from 0° to 196°) by repeating the geometric calibration (not interpolation) 4 times over an 8-hour period of normal use. Three scenarios were evaluated: (i) a conventional pre-calibrated circular orbit (denoted as "Calibrated Circular"); (ii) a pre-determined and pre-calibrated non-circular orbit (denoted as "Calibrated Non-Circular"); and (iii) a general non-circular orbit for which the projection matrix is determined by the interpolation-based method described above (denoted as "Interpolated Non-Circular"). Basic image quality characteristics of scans under these three scenarios were evaluated in terms of spatial resolution (modulation transfer function, MTF) and 3D sampling characteristics (cone-beam artifacts) in a head phantom containing a variety of custom inserts. The axial plane MTF was determined from the edge-spread function measured from a high-contrast (~300 HU) cylindrical rod insert. The magnitude of cone-beam artifacts was measured in terms of the full-width-at-half-maximum (FWHM) of the superior-inferior edges of flat disks ("Defrise phantom") inserted within the cranium. All scans involved 400 projections over a 196° arc in $\theta$ at 110 kV, 1.65 mAs / view, with a 30 s scan time.

### 2) Metal Artifact Avoidance (MAA) with Non-Circular Orbits

The performance of the CNN-based metal object localization step was evaluated in terms of Dice coefficient in the validation dataset and in the testing dataset [scans of a cadaver instrumented with six pairs of pedicle screws (Nuvasive, San Diego USA)]. We compared the performance of the proposed end-to-end method with two other methods: (i) conventional image domain U-Net segmentation on direct backprojection of scout views (referred to as the "single U-Net");[5] and (ii) projection domain U-Net segmentation of metal objects, followed by backprojection and image domain U-Net segmentation (referred to as the "dual U-Net"). Note that the two U-Nets in (ii) were trained separately (not end-to-end).

The performance of the MAA method was evaluated in a phantom study involving an anthropomorphic chest phantom containing a natural human skeleton in tissue-equivalent plastic and 8 spinal pedicle screws (DePuy-Synthes, Raynham USA; ranging 30-60 mm in length). The screws were placed with out-of-plane angle ranging from -20° to +30° (positive on one side of the spine, negative on the other). Metal artifact magnitude was assessed in terms of "blooming" about the shaft of the screw (FWHM of the screw minus its true width) for conventional circular orbit scans and the optimal non-circular orbit defined by MAA.

# III. RESULTS

## A. Geometric Calibration of Non-Circular Orbits

Figure 5 shows the degree of reproducibility in system geometry for the pre-determined non-circular orbit described in §C.1. A reduction in reproducibility was observed: for example, the standard deviation in piercing point location $(u_0, v_0)$ over repeat trials was ~1.7 mm (vs. ~0.8 mm for a conventional circular orbit). Note that the mobile C-arm was not electromechanically tuned beyond its standard clinical deployment, which does *not* support non-circular orbits in standard use. As shown below, however, despite the reduced mechanical reproducibility, the interpolation-based calibration provided a reasonable estimate of system geometry supporting 3D image reconstruction.

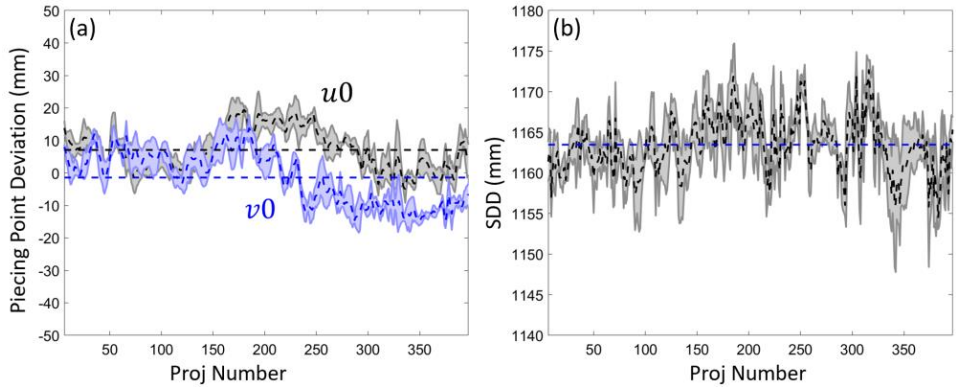

Fig. 5. Reproducibility of geometric parameters for the pre-determined non-circular orbit described in §C.1. (a) Location of the piecing point $(u_0, v_0)$ and (b) SDD in 4 repeat trials over an 8-hour interval of routine use.

Figure 6 illustrates how errors in geometric calibration relate to reduction in spatial resolution for the three scenarios defined in §C.1. Compared to the standard "Calibrated Circular" scenario, the "Calibrated Non-Circular" scenario showed minor reduction in MTF (~3.5% reduction in the spatial frequency at which MTF = 0.10, denoted as $f_{10}$) due to the decreased reproducibility in calibration parameters. The "Interpolated Non-Circular" scenario showed further reduction in resolution (~12% reduction in $f_{10}$) due to two effects: smoothing of geometric parameters in the interpolation; and ignoring differences in gantry momentum for a continuous non-circular orbit compared to the discrete library of circular scans. Despite the reduction in MTF, images obtained with the "Interpolated Non-Circular" scenario appear visually comparable to the others, as in Fig. 6(b-c). Therefore, the interpolation-based method appears to provide a reasonable estimate of system geometry for non-circular scans for which neither a pre-calibration nor a "self-calibration" (using previous scan and 3D-2D registration [14]) is available.

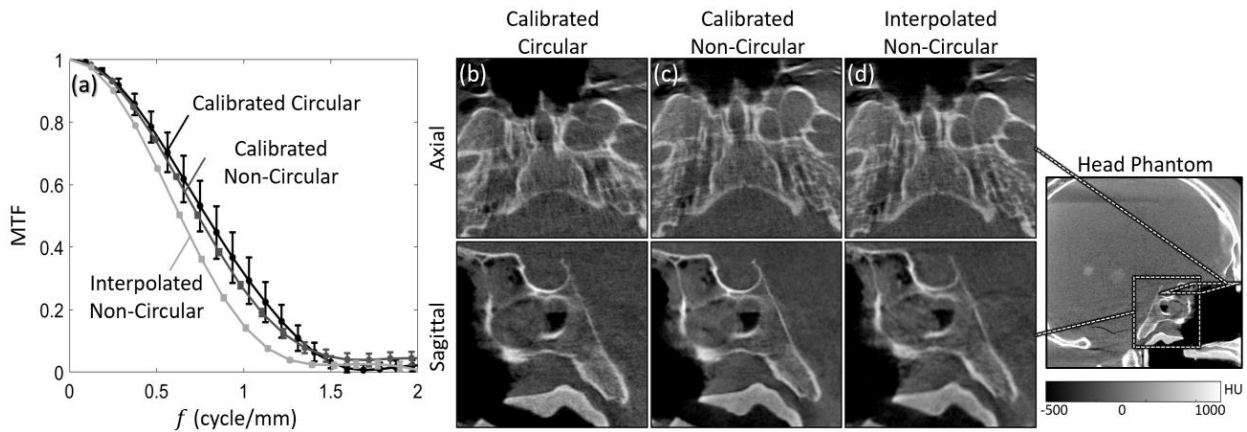

Fig. 6. Effect of geometric calibration accuracy on spatial resolution (MTF) and 3D image quality. (a) MTF for the three scenarios defined in §C.1. (b-d) Axial and sagittal zoomed-in views (skull-base) of a head phantom for the three scenarios.

Figure 7 shows the expected reduction in cone-beam artifacts from non-circular orbits in sagittal images of a head phantom containing stacks of flat disks (§C.1). The apparent thickness (FWHM) of the uppermost disk was reduced from 9 mm for the standard "Circular Protocol" to its true thickness (~5.5 mm) for both of the Non-Circular scenarios.

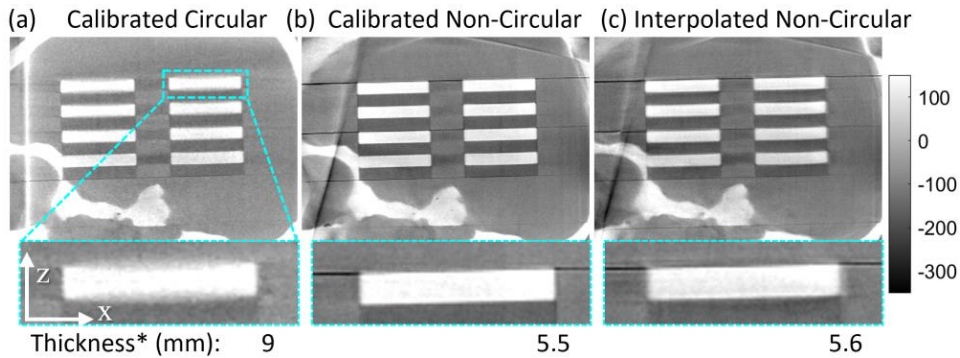

Fig. 7. Reduction of cone-beam artifacts for the non-circular orbits (§C.1) evident in sagittal images of a head phantom containing flat disk inserts.

### B. MAA with Non-Circular Orbits

Figures 8 and 9 show the segmentation performance of the end-to-end network to outperform the single and dual U-Nets. The increase in Dice coefficient (especially for fewer scout views) confirms the effectiveness of end-to-end training: for just two scout views, the end-to-end method increased Dice by ~29% in the validation dataset, and by ~42% in the testing dataset compared to the dual U-Net. This is consistent with other works in literature that utilizes end-to-end training.[17], [18] Figure 8(b) shows that ≥8 filters in the projection domain ($N_c$) were needed for reasonable performance of the end-to-end network. Figure 9 shows that with only two scout views, the end-to-end network was able to capture the shape and direction of all six pairs of bilateral screws in the testing dataset (real cadaver data, unlike simulated data used as validation dataset). While the Dice coefficient (0.7) would still be considered low for some applications (e.g., MAR) and was shown to improve with more scout views, it is more than sufficient for MAA orbit optimization, as shown below.

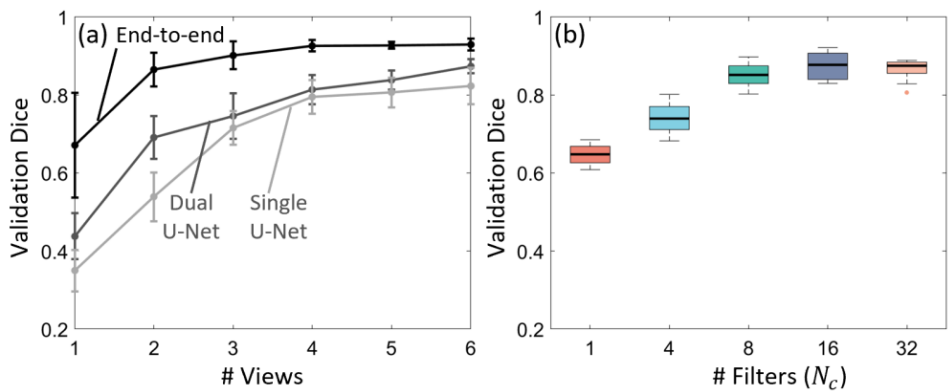

Fig. 8. Performance of CNN in 3D localization of metal instruments from sparse projection views for the validation dataset. (a) Validation Dice for the three methods described in §C.2 evaluated as a function of the number of scout views. (b) Validation Dice for the end-to-end network with respect to the number of filters in the projection domain ($N_c$, as shown in Fig. 3).

Figure 10 shows the $q_{poly}(\theta, \phi)$ metric map computed by the MAA method with just two scout views of the chest phantom and spine screws. Clearly, there is no circular orbit [horizontal line in (a)] that would substantially reduce metal artifact for all screws, as there is always at least one region of strong metal-induced bias in the $(\theta, \phi)$ trajectory space (roughly corresponds to the out-of-plane angle of the screw). This issue is resolved by 2D minimization of Eqs. (6-7), resulting in the non-circular orbit marked by the cyan curve (a). The non-circular orbit avoids most of the low-fidelity views, steering a path in $(\theta, \phi)$ that minimizes variation in spectral shift. The resulting image quality is shown in Figs. 10(c)-(f), showing strong reduction in blooming artifacts about the screw (~70% improvement in screw shaft blooming achieved with the optimal non-circular orbit).

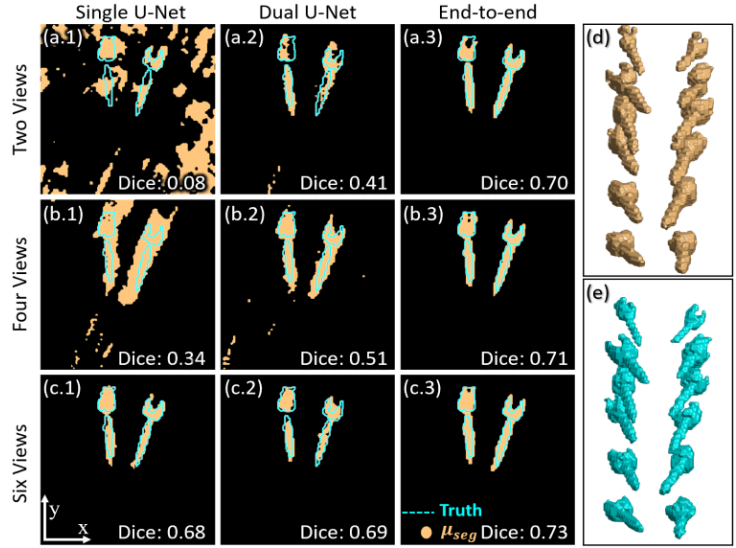

Fig. 9. Segmentation performance of three neural networks in the testing dataset (cadaver with 12 pedicle screws). (a)-(c) Example axial slice segmentation overlaid with ground truth (cyan). Segmentations are shown for varying number of scout views. (d) Isosurface of $b_{seg}$ computed from two views for the end-to-end method. (e) Isosurface of ground truth segmentation (downsampled the same as $b_{seg}$).

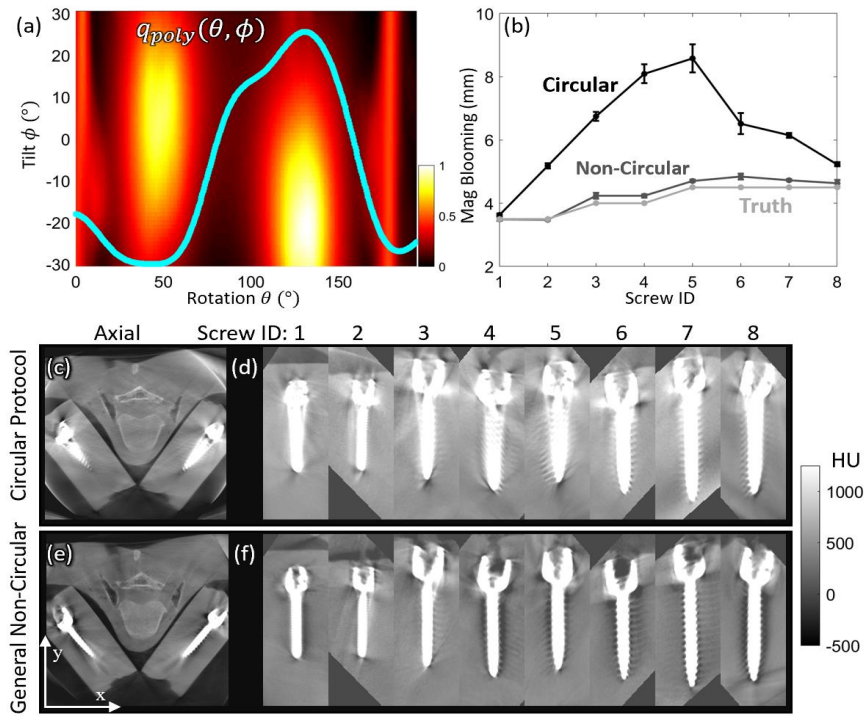

Fig. 10. The MAA method applied to a chest phantom implanted with eight pedicle screws. (a) The $q_{poly}(\theta, \phi)$ metric map overlaid with the optimal non-circular orbit (cyan). (b) Magnitude of blooming artifact (FWHM of the screw shaft) for each of the 8 screws. Axial images in (c and e) for circular and (MAA) non-circular scans show the improvement in visual image quality acquisition, illustrated further in (d and f) by zoomed quasi-axial slices in-plane with each screw.

## IV. CONCLUSION

A method for geometric calibration of non-circular C-arm CBCT orbits was demonstrated, giving a practical means to estimate system geometry from a discrete library of circular scan calibrations. Although the geometric calibration method carried a measurable reduction in MTF, the effect on visual image quality was relatively minor. CBCT images acquired from non-circular orbits were shown to improve 3D sampling characteristics (e.g., reduction of cone-beam artifacts), as expected. Such capability enabled a method (called MAA) that identifies non-circular orbits with minimal metal-induced biases. A novel end-to-end neural network was shown to localize metal objects from just two scout views without strong prior information of the patient anatomy or metal instruments. Integration of the end-to-end network with the MAA method for non-circular orbits demonstrated strong reduction in metal artifacts in phantom and cadaver studies. Moreover, the method is compatible with established MAR and polyenergetic reconstruction algorithms to further reduce artifacts. Ongoing / future work includes adding the attention gate mechanism into the end-to-end network and improved geometric calibration method for general non-circular orbits.

## ACKNOWLEDGEMENT

The research was supported by NIH R01-EB-017226 and academic-industry partnership with Siemens Healthineers. This work was presented at the 6th International Conference on Image Formation in X-Ray Computed Tomography, August, 2020, Regensburg, Germany.